\documentclass[twocolumn,showpacs,preprintnumbers,amsmath,amssymb]{revtex4}

\usepackage{float}
\usepackage{graphicx}
\usepackage{dcolumn}
\usepackage{bm}

\begin{document}

\title{Coupled-resonator-induced transparency with a squeezed vacuum}
\author{Ke Di }
\author{Changde Xie }
\author{Jing Zhang}
\email{jzhang74@yahoo.com,jzhang74@sxu.edu.cn}

\affiliation{ State Key Laboratory of Quantum Optics and Quantum
Optics Devices, Institute of Opto-Electronics, Shanxi University,
Taiyuan 030006, P.R.China }

\begin{abstract}
We present the first experimental observation of quantum fluctuation
spectra in two coupled optical cavities with an injected squeezed
vacuum light. The quadrature components of the reflected squeezed
vacuum spectra are measured by phase sensitive homodyne detector.
The experimental results demonstrate coupled-resonator-induced
transparency in the quantum regime, in which
electromagnetically-induced-transparency-like characteristic of the
absorption and dispersion properties of the coupled optical cavities
determines the line-shape of the reflected quantum noise spectra.
\end{abstract}

\maketitle

The squeezed vacuum as a vital source of quantum entanglement enable
us to implement a variety of quantum information protocols
deterministically \cite{eight,nine}. Propagation, storage, and
manipulation of the squeezed vacuum and entanglement are essential
for quantum information processing \cite{eight,nine}. Recent
experiments have demonstrated the transmission \cite{one}, slowing
down \cite{two,three,four}, storage and retrieval \cite{five,six} of
squeezed states of light through electromagnetically induced
transparency (EIT) in multi-level atomic systems \cite{seven}. The
manipulations of squeezed vacuum and two-mode entangled state by
means of a phase-sensitive optical amplifier \cite{seven1} have also
been achieved experimentally.

EIT in atoms comes from quantum destructive interferences between
excitation pathways to the upper level of an atomic three-level
system \cite{seven}. It has been proved that there are analogical
phenomena of the coherence and interference process in some
classical physical systems, such as plasma \cite{ten}, coupled
optical resonators \cite{eleven,eleven0,eleven1}, mechanical or
electric oscillators \cite{twelve}, and optical parametric
oscillators \cite{thirteen}. Particularly, the analog of EIT in
coupled optical resonators including the capabilities of slowing,
stopping, storing and time reversing an incident optical pulse have
made great progress in experiment recently, which have been reported
for observing the structure of the EIT-like spectrum in a compound
glass waveguide platform using relatively large resonators
\cite{forteen}, coupled fused-silica microspheres
\cite{fifteen,sixteen}, integrated micron-size silicon optical
resonator systems \cite{seventeen}, photonic crystal cavities
\cite{eighteen}, and fiber ring resonators \cite{nineteen}. These
works also open up the new possibility of utilizing coupled optical
resonators to realize the optical communication and the simulation
of coherent effect in quantum optics. In this Letter, we report the
first experimental observation of quantum fluctuation spectra in two
coupled optical cavities with an injected squeezed vacuum state.
Using the phase sensitive homodyne detection setup with a local
oscillator beam, the line-shapes of the quantum noise spectra of the
output squeezed vacuum are experimentally measured, which clearly
exhibit the EIT-like characteristic of the absorptive and dispersive
properties in two coupled optical cavities. We are sure that the
results demonstrated in the experiment using coupled two large
optical cavities can be obtained identically in micro-structure
devices if on-chip coupled optical cavities are used, and thus it
has potential applications in the manipulation of optical quantum
states.

\begin{figure}[H]
\centering{
\includegraphics[width=3.5in]{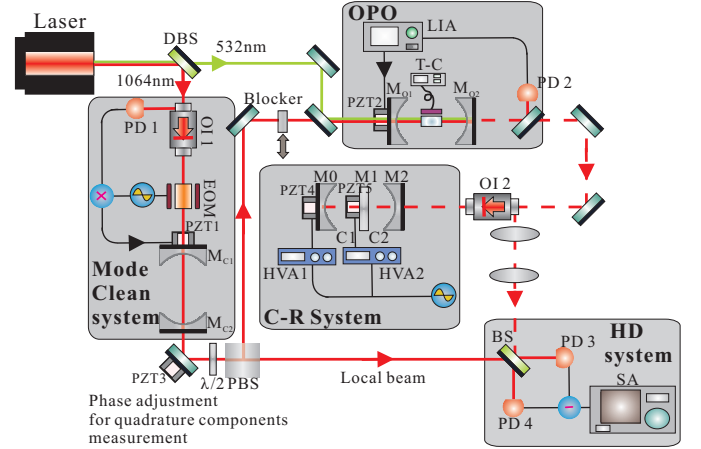}
} \vspace{0.1in}

\caption{\label{Fig:epsart} (Color online) Schematic of the
experimental setup. OPO: an optical parametric oscillation below
threshold to generate the squeezed vacuum state; DBS: Dichroic beam
splitter; PD: photodiode; OI: optical isolator; EOM: electro-optical
modulator; T-C: temperature controller; LIA: Lock-in amplifier; HVA:
high voltage amplifier; PBS: polarizing beam splitter; SA: spectrum
analyzer; HD: homodyne detector; PZT: piezoelectric transducer.}
\end{figure}

The experimental setup (Fig. 1) consists of five parts including a
laser source, a mode clean system, an optical parametric oscillation
(OPO), coupled optical resonators (C-R system), and homodyne
detection (HD) system. The laser source is a diode-pumped
intracavity frequency-doubled (continuous-wave ring
Nd:YVO$_{4}$+KTP) laser, which provides the second-harmonic light of
$\sim$ 200 $mW$ at 532 $nm$ and the fundamental light of 50 $mW$ at
1064 $nm$ simultaneously. The fundamental light is injected into a
mode clean system which is served as spatial mode cleaner. The
fundamental light is divided into two optical beams after it passes
through the mode clean cavity, one of which is used for the local
oscillator in the HD system, and the other one is sent into the OPO
to be as the signal field for the alignment of both OPO cavity and
the C-R system. When this signal beam injected into OPO is blocked
by a movable blocker and only the pump light at 532 $nm$ exists, the
output of OPO will be a squeezed vacuum. The configuration of the
OPO is a near concentric optical cavity consisting of two concave
mirrors $M_{o1}$ and $M_{o2}$ of 30 $mm$ curvature radius with a
separated distance of $\sim60$ $mm$ and a 12 $mm$ long PPKTP
(periodically poled KTP) crystal is place at the center position of
the cavity. The PPKTP crystal is mounted in a copper block for
actively controlling its temperature at the phase-matching
temperature. The input coupler $M_{o1}$ of the OPO cavity has a
reflectivity of 99.5$\%$ at 1064 $nm$ and 30$\%$ at 532 $nm$, which
is mounted on a piezoelectric transducer (PZT2) for the adjustment
of the cavity length. The reflectivity of the output coupler
$M_{o2}$ is 88.7$\%$ at 1064 $nm$ and it is coated to be a high
reflector ( $>$99$\%$) for 532 $nm$. In such a coated optical cavity
the power of the second-harmonic  wave can be built up to a few
times of the input power when the fundamental wavelength is on
resonance with the cavity. The squeezed vacuum generated by OPO is
injected into the coupled resonators through an optical isolator,
which separates the reflection field of the coupled resonators from
the injected field. The squeezing degree of the squeezed vacuum
generated by the OPO is measured by a balanced homodyne detector
consisting of a 50$\%$ beam splitter, where the squeezed vacuum and
the local beam are combined, as well as a pair of Epitaxx ETX-500T
photodiodes. Adjusting the spatial mode of reflected beam from the
coupled cavities, an interference fringe visibility of 94$\%$
between local oscillator and the reflected beam was achieved in the
experiment.

The C-R system in Fig. 1 involves the two coupled resonators
consisting of two standing wave cavities, which are constructed by
M0, M1, and M2. M0 and M2 are the concave mirrors of 30 $mm$
curvature radius with a separation of 59 $mm$. M1 is a plane mirror
mounted on a PZT5, which is inserted at the middle between M0 and
M2. One surface of M1 facing to M2 is coated with anti-reflection
for 1064 $nm$ and the other surface facing to M0 is coated with
partial reflectivity for 1064 $nm$. The two directly coupled optical
cavities (C1 and C2) are in the same geometric structure, and the
coupling strength between C1 and C2 depends on the reflectivity of
the middle mirror M1. The PZT 4 and 5 mounted on the mirrors M0 and
M1 are used for controlling the cavity length of C1 and C2,
respectively. The reflectivity of M2 is 96.8$\%$ and M0 is high
reflectivity for 1064 $nm$. When the reflectivity of M1 is chosen to
be $99.8\%$ (case I) and $96.7$ (case II) respectively, two kinds of
coupled resonators with different coupling strength are constructed.

\begin{figure}[H]
\centering{
\includegraphics[width=3.5in]{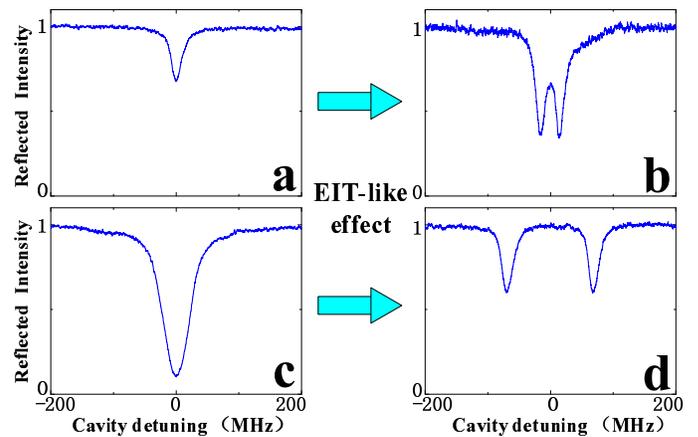}
} \vspace{0.1in}

\caption{\label{Fig:epsart} (Color online) The reflection intensity
of the coupled cavities injected with the coherent signal beam, as a
function of the cavity detuning for different coupling strength. (a)
and (c) correspond to the reflected intensity for the single cavity
C2 (blocking the cavity C1). (b) and (d) are the reflected intensity
of the two coupled resonators with two different coupling strength
of the case I and II respectively. }
\end{figure}

Before injecting the squeezed state into the two coupled cavities,
we first characterize the two coupled cavities in the classical
field by injecting a coherent signal beam at the subharmonic
wavelength. When the pump beam is blocked, the OPO cavity is locked
on resonance with the fundamental field by a lock-in amplifier and
the transmitted light at 1064 $nm$ is injected into the coupled
resonator system to be the signal beam. The intensity of the
reflected field of the coupled resonators extracted by an optical
isolator is detected directly by a photodiode. When we blocked the
cavity C1 of the coupled cavities, the reflected intensity from the
cavity C2 during scanning the cavity length are a simple Lorentzian
profile as shown in the Fig. 2(a) and (c). Fig. 2(a) and (c)
correspond to the over-coupled optical cavity for the experimental
parameters (case I) and the near critically-coupled optical cavity
(case II), respectively. Then we open the cavity C1 to study the
features of the two coupled optical cavities. Usually the reflected
intensity with the injected coherent signal light are measured by
adjusting two cavities to be co-resonant and scanning the laser
frequency \cite{fifteen,sixteen,seventeen}. Here we employ an
equivalent way to measure the intensity of the reflected light, that
is to keep the laser frequency unchanged and scan the length of the
two cavities. A high voltage amplifier (HVA1), which that drives the
PZT4 actuating the mirror M0 uses the same scanning signal source
with the HVA2 driving the PZT5 for the mirror M1. So, when the gain
of HVA1 is two times larger than that of HVA2, the reflected
intensity by scanning the length of two cavities are equivalent to
scanning the laser frequency. As we can see in the Fig. 2(b), a
transparent window appeared in the middle of the absorption profile
due to the existence of the cavity C2. The destructive interference
between the two optical pathways results in the transparent peak,
which is named as a coupled-resonator-induced transparency. This
transparent peak becomes broader when the coupling strength is
increased and the original mode eventually is split into two modes
(as shown in Fig. 2(d)).

Now, let's inject the squeezed vacuum state into the coupled optical
cavitied system to investigate the coupled-resonator-induced
transparency effect formed by the quantum field. A quadrature
squeezed vacuum state with about 1.6 $dB$ squeezing component and 4
$dB$ antisqueezing component at the sideband frequency of 2.5 $MHz$
is generated from the subthreshold OPO when we block the signal beam
of 1064 $nm$ light and only approximate 50 $mw$ pump beam of 532
$nm$ is applied. Similarly, the reflection spectra of the squeezed
vacuum light from the cavity C2 is examined by scanning M1 with PZT5
when blocking the cavity C1 of the coupled cavities, firstly. Fixing
the relative phase between the local beam and the squeezed vacuum
light reflected by the cavity C2 to be $\theta=0$ (or $\theta=\pi/2$
by tuning the PZT3, we can obtain the squeezing (or antisqueezing)
component of the reflected quadrature spectra at far off resonance.
Because of the absorption and dispersion properties of the single
optical cavity, the reflected spectra of quantum fluctuation near
resonance will be changed \cite{twenty1}. Fig. 3(a) and Fig. 4(a)
show the reflected quadrature spectra of the squeezing component for
the experimental parameters of case I and II respectively, and Fig.
3(b) and Fig. 4(b) show the corresponding antisqueezing component.
The lineshape of the reflected spectrum of the squeezed component
for the over-coupled optical cavity (case I) presents M profile
(Fig. 3(a)), and the two shoulders at detuning frequencies are
higher, which is due to the lower loss and larger phase shift in the
over-coupled cavity. The two shoulders in the profile corresponds to
the phase shift of $\pm\pi/2$ induced by the cavity. The degree of
the squeezing at zero detuning is below the SNL and will reach the
degree of the input squeezing for the strongly over-coupled cavity.
The noise spectrum of the antisqueezed component presents the W
profile as shown in Fig. 3(b). The noise level at zero detuning will
reach the level of the input antisqueezing component for the
strongly over-coupled cavity. Since the amplitude of the reflected
field is zero at resonance for critically-coupled cavity (case II),
the lineshape of the reflected spectrum of the squeezed component
also presents M profile (Fig. 4(a)), however the two shoulders are
lower than that of case I. The reflected spectrum of the
antisqueezing component presents V profile and only one dip locates
at resonance (Fig. 4(b)).

\begin{figure}[H]
\centering{
\includegraphics[width=3.5in]{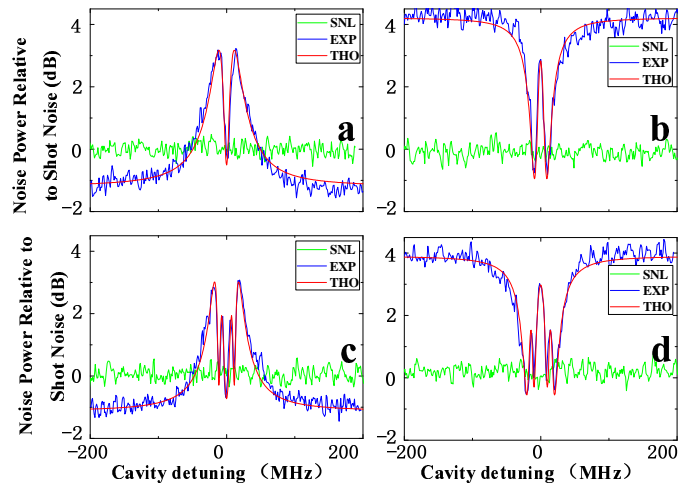}
} \vspace{0.1in}

\caption{\label{Fig:epsart} (Color online). The quantum fluctuation
spectra of the reflected field from the coupled optical cavities for
the experimental parameter case I. The quadrature spectra are
measured at the sideband frequency of 2.5 $MH$z. (a) and (b) show
respectively the reflected quadrature spectra of the squeezing and
antisqueezing component for the single cavity C2. (c) and (d) are
the reflected quadrature spectra of the squeezing and antisqueezing
component respectively for two coupled resonators. The red curves
are the theoretical calculations. The green curves are the
shot-noise limit. The blue curves are the experimental results.}
\end{figure}

Now, we open the cavity C1 to build up the coupled resonators
system. The reflection spectra for the squeezed vacuum light are
studied by the way of scanning the length of two cavities as
described above. Fig. 3(c), Fig. 4(c) and Fig. 3(d), Fig. 4(d) show
the reflected quadrature spectra of the squeezing and antisqeezing
components for the experimental parameters of case I and II
respectively. For two coupled resonators with weak coupling
strength, two extra dips appear respectively in the two shoulders of
the reflected spectrum for the squeezed component (three dips appear
in the center of a broader peak as shown in Fig. 3(c)).
Correspondingly, three peaks present in the center of a broader dip
for the reflected spectrum of the antisqueezing component (Fig.
3(d)). These extra dips and peaks are induced by the dispersion
characteristic of the coupled-resonator-induced transparency, which
presents three points of crossing the zero phase shift. For the case
of large coupling strength, in which the original mode splits into
two, two independent M profiles (each M profile shows the same
lineshape of the reflected spectrum of the squeezed component for
the single over-coupled cavity) appear in the detuning frequencies
of the reflected spectrum of the squeezed component as shown in Fig.
4(c). Correspondingly, two independent W profiles also appear in the
detuning frequencies of the reflected spectrum of the antisqueezed
component as shown in Fig. 4(d). In order to make theoretical
comparisons with the above experimental results, the reflected
spectra for the amplitude and phase quadratures are calculated (see
Ref. \cite{twenty-one} for the detailed theoretical calculation)
with experimental parameters. Our theoretically calculated results
(red lines) are plotted together with the experimental data in Figs.
3 and 4, which show excellent agreements. This result shows that
EIT-like characteristic of the absorption and dispersion properties
of the coupled optical cavities determines the line-shape of the
reflected quantum noise spectra.

\begin{figure}[H]
\centering{
\includegraphics[width=3.5in]{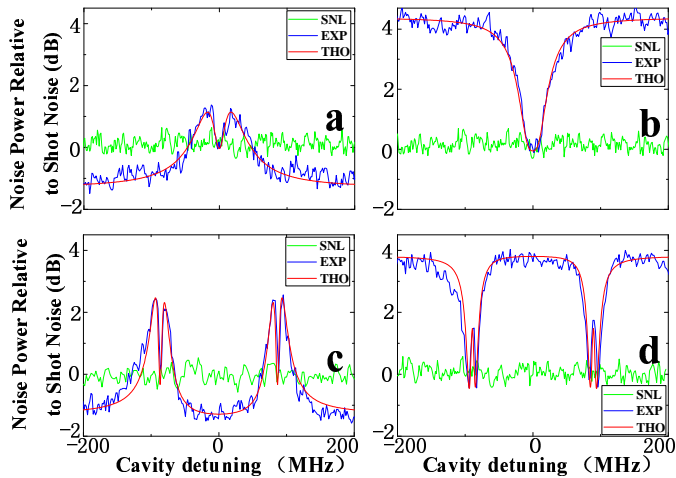}
} \vspace{0.1in}

\caption{\label{Fig:epsart} (Color online). The quantum fluctuation
spectra of the reflected field from the coupled optical cavities for
the experimental parameter case II. (a) and (b) are the reflected
quadrature spectra of the squeezing and antisqueezing component for
the single cavity C2, respectively. (c) and (d) are the reflected
quadrature spectra of the squeezing and antisqueezing component for
two coupled resonators, respectively.}
\end{figure}

In summary, we have experimentally demonstrated the response of two
coupled optical cavities in quantum domain. We developed a method to
measure the reflection spectra by scanning the length of two
cavities instead of scanning the laser frequency. We gave the
reflected quantum fluctuation spectra of the two coupled resonators
with two different coupling strength and showed that the line-shapes
of the reflected quantum noise spectra depend on the absorptive and
dispersive properties of two coupled optical cavities with EIT-like
characteristic. This work achieves the first step of manipulating
quantum fluctuations using coupled resonators and provides a scheme
for the future studies on slow light, storage and retrieval
\cite{eleven1} of quantum fields. Here we would like to emphasize
that the basic requirements of a light-stopping process (capture,
storage and release of the light pulse) are that the
coupled-resonator system supports a large-bandwidth state to
accommodate the input pulse bandwidth, which is then dynamically
tuned to an narrow-bandwidth state to stop the pulse and done
reversibly after some storage time to release light pulse. Two
dynamically tuned resonators for stopping light have been proposed
theoretically \cite{eleven0,twenty-two} and demonstrated
experimentally \cite{twenty-three}. Thus this protocol should be
feasible to be used to capture, storage and release of squeezed
light. The current structure of two coupled cavities in our
experiment can not satisfy this requirement completely to tune the
cavity bandwidth dynamically. However, the structure with two
dynamically tuned resonators above-mentioned for stopping light
could be used in our system. We also hope that this work will
stimulate the development of manipulating the quantum optical fields
with on-chip coupled optical cavities.

\smallskip \acknowledgments
This research was supported in part by NSFC for Distinguished Young
Scholars (Grant No. 10725416), National Basic Research Program of
China (Grant No. 2011CB921601,2010CB923103), NSFC Project for
Excellent Research Team (Grant No. 60821004), and NSFC (Grant No.
60736040).

\maketitle{ \textbf{ SUPPLEMENTARY MATERIAL}}

In this Appendix we describe in detail theoretical calculations of
the quantum fluctuation spectra of the reflected field from the
coupled optical cavities. The complex electric-field transmission
($E_{out}/E_{in}$) from two coupled optical cavities is given by
\cite{one,two}
\begin{eqnarray}
R_{2}(\phi_{2},\phi_{1})=\frac{r_{2}-R_{1}a_{2}\exp(i\phi_{2})}{1-r_{2}R_{1}a_{2}\exp(i\phi_{2})},
\end{eqnarray}
where $R_{1}$ is the complex reflectivity from the first cavity C1
\begin{eqnarray}
R_{1}(\phi_{1})=\frac{r_{1}-a_{1}\exp(i\phi_{1})}{1-r_{1}r_{0}a_{1}\exp(i\phi_{1})}.
\end{eqnarray}
Here, $\phi_{j}=2\pi 2L_{j}n_{j}\omega/c$ are the round trip
phase-shifts, $n_{j}$ are the refractive indices, $L_{j}$ are the
cavity length, $\omega$ is the frequency of the laser, and $j=1,2$
specifies the the first or second cavity. $r_{j}$ are the reflection
indices of cavity mirrors, and $a_{j}$ are the round trip
attenuation coefficients. The reflected field given in Eq. (1) can
be separated into the amplitude and phase of the coupled cavities
\begin{eqnarray}
\rho(\phi_{2},\phi_{1})=Abs(R_{2})\nonumber\\
\theta(\phi_{2},\phi_{1})=Arg(R_{2})
\end{eqnarray}

The quantum states we consider in this paper are described with the
electromagnetic field annihilation operator $
\hat{a}=(\hat{X}+i\hat{Y})/2$, which is expressed in terms of the
amplitude $\hat{X}$ and phase $\hat{Y}$ quadrature with the
canonical commutation relation $[\hat{X},\hat{Y}]=2i$. Since the
quantum fluctuation spectra of the quadrature variances of the
reflection field are measured and analyzed, we implement the Fourier
transformation $\hat{a}(\Omega)=\frac{1}{\sqrt{2\pi}}\int
dt\hat{a}(t)e^{-i\Omega t}$. The amplitude and phase quadratures of
the reflected field in frequency domain are expressed by
\begin{eqnarray}
\hat{X}_{out}(\Omega)&=&\hat{a}_{out}(\Omega)+\hat{a}_{out}^{\dag}(-\Omega),\nonumber\\
\hat{Y}_{out}(\Omega)&=&-i[\hat{a}_{out}(\Omega)-\hat{a}_{out}^{\dag}(-\Omega)].
\end{eqnarray}

The relationship of the input and reflected quantum fields from two
coupled optical cavities is calculated based Eqs. (1) and (3)
\begin{eqnarray}
\hat{a}_{out}(\Omega)&=&\rho(\omega_{0}+\Omega)\exp[i\theta(\omega_{0}+\Omega)]\hat{a}_{in}(\Omega)\nonumber\\
&&+\sqrt{1-\rho^{2}(\omega_{0}+\Omega)}\hat{a}_{v}(\Omega),\nonumber\\
\hat{a}^{\dag}_{out}(-\Omega)&=&
\rho(\omega_{0}-\Omega)\exp[-i\theta(\omega_{0}-\Omega)]\hat{a}^{\dag}_{in}(-\Omega)\nonumber\\
&&+\sqrt{1-\rho^{2}(\omega_{0}-\Omega)}\hat{a}^{\dag}_{v}(-\Omega),
\end{eqnarray}
where $\hat{a}_{v}$ is the vacuum field coupled into the vacuum
field due to the loss, $\omega_{0}$ is central frequency of quantum
field (belong to optical frequency) and $\Omega$ is the analysis
frequency of the sidebands. Thus we may obtain the amplitude
quadrature and its variance of the reflected field
\begin{eqnarray}
\hat{X}_{out}(\Omega)&=&\hat{a}_{out}(\Omega)+
\hat{a}_{out}^{\dag}(-\Omega)\nonumber\\
&=&\rho(\omega_{0}+\Omega)\exp[i\theta(\omega_{0}+\Omega)]\nonumber\\
&&\times\frac{\hat{X}_{in}(\Omega)+i\hat{Y}_{in}(\Omega)}{2}\nonumber\\
&&+\sqrt{1-\rho^{2}(\omega_{0}+\Omega)}\frac{\hat{X}_{v}(\Omega)+i\hat{Y}_{v}(\Omega)}{2}\nonumber\\
&&+\rho(\omega_{0}-\Omega)\exp[-i\theta(\omega_{0}-\Omega)]\nonumber\\
&&\times\frac{\hat{X}_{in}(\Omega)-i\hat{Y}_{in}(\Omega)}{2}\nonumber\\
&&+\sqrt{1-\rho^{2}(\omega_{0}-\Omega)}\frac{\hat{X}_{v}(\Omega)-i\hat{Y}_{v}(\Omega)}{2},\label{x} \\
\langle\delta^{2}\hat{X}_{out}(\Omega)\rangle&=&\frac{1}{4}|\rho(\omega_{0}+\Omega)\exp(i\theta(\omega_{0}+\Omega))\nonumber\\
&&+\rho(\omega_{0}-\Omega)\exp(-i\theta(\omega_{0}-\Omega))|^{2}\langle\delta^{2}\hat{X}_{in}(\Omega)\rangle\nonumber\\
&&+\frac{1}{4}|\rho(\omega_{0}+\Omega)\exp(i\theta(\omega_{0}+\Omega))\nonumber\\
&&-\rho(\omega_{0}-\Omega)\exp(-i\theta(\omega_{0}-\Omega))|^{2}\langle\delta^{2}\hat{Y}_{in}(\Omega)\rangle\nonumber\\
&&+\frac{1}{4}|\sqrt{1-\rho^{2}(\omega_{0}+\Omega)}\nonumber\\
&&+\sqrt{1-\rho^{2}(\omega_{0}-\Omega)}|^2\langle\delta^{2}\hat{X}_{v}(\Omega)\rangle\nonumber\\
&&+\frac{1}{4}|\sqrt{1-\rho^{2}(\omega_{0}+\Omega)}\nonumber\\
&&-\sqrt{1-\rho^{2}(\omega_{0}-\Omega)}|^2\langle\delta^{2}\hat{Y}_{v}(\Omega)\rangle.
\end{eqnarray}
Similarly, we get the variance of the phase quadrature of the
reflected field
\begin{eqnarray}
\langle\delta^{2}\hat{Y}_{out}(\Omega)\rangle&=&\frac{1}{4}|-\rho(\omega_{0}+\Omega)\exp(i\theta(\omega_{0}+\Omega))\nonumber\\
&&+\rho(\omega_{0}-\Omega)\exp(-i\theta(\omega_{0}-\Omega))|^{2}\langle\delta^{2}\hat{X}_{in}(\Omega)\rangle\nonumber\\
&&+\frac{1}{4}|\rho(\omega_{0}+\Omega)\exp(i\theta(\omega_{0}+\Omega))\nonumber\\
&&+\rho(\omega_{0}-\Omega)\exp(-i\theta(\omega_{0}-\Omega))|^{2}\langle\delta^{2}\hat{Y}_{in}(\Omega)\rangle\nonumber\\
&&+\frac{1}{4}|-\sqrt{1-\rho^{2}(\omega_{0}+\Omega)}\nonumber\\
&&+\sqrt{1-\rho^{2}(\omega_{0}-\Omega)}|^2\langle\delta^{2}\hat{X}_{v}(\Omega)\rangle\nonumber\\
&&+\frac{1}{4}|\sqrt{1-\rho^{2}(\omega_{0}+\Omega)}\nonumber\\
&&+\sqrt{1-\rho^{2}(\omega_{0}-\Omega)}|^2\langle\delta^{2}\hat{Y}_{v}(\Omega)\rangle.
\end{eqnarray}
Here, the variances of the vacuum field are normalized
$\langle\delta^{2}\hat{X}_{v}(\Omega)\rangle=\langle\delta^{2}\hat{Y}_{v}(\Omega)\rangle=1$.
The variances of the input squeezed vacuum field with squeezing
factor $s$ can be expressed by
$\langle\delta^{2}\hat{X}_{in}(\Omega)\rangle=\exp(-2s)$, and
$\langle\delta^{2}\hat{Y}_{in}(\Omega)\rangle=\exp(2s)$.

\begin{figure}[H]
\centering{
\includegraphics[width=3.5in]{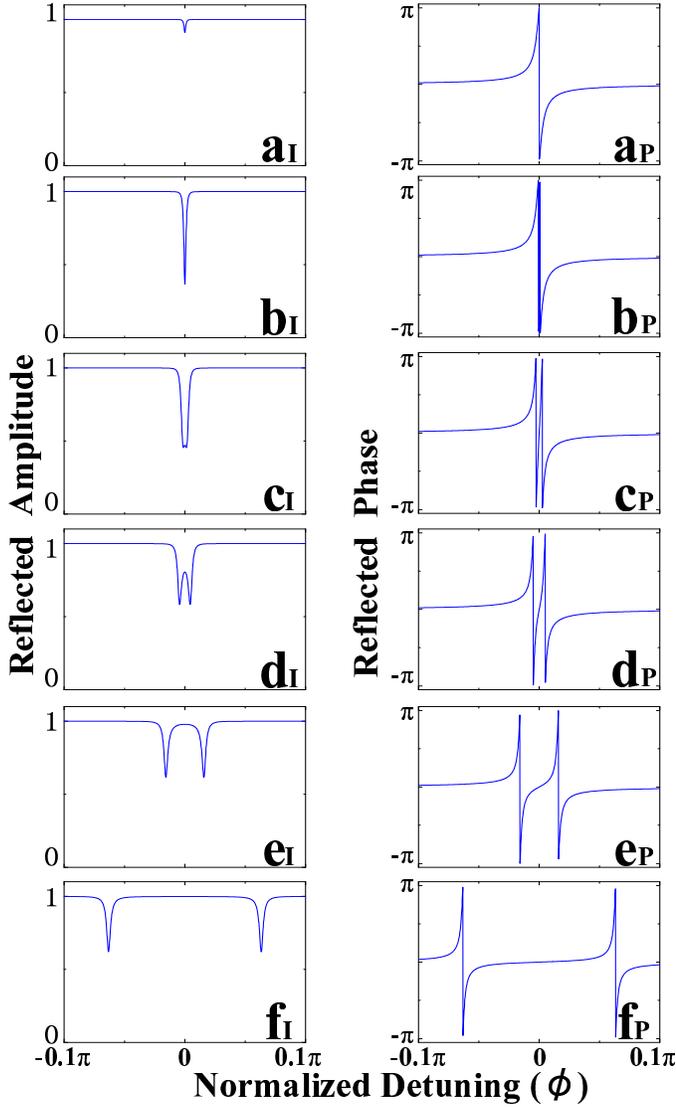}
} \vspace{0.1in} \caption{\label{Fig:epsart} (Color online).
Theoretical plots of the reflected amplitude and phase of the two
coupled cavities as a function of the cavity detuning for different
coupling strength. $\mathrm{a_{I}-f_{I}}$ the reflected amplitude of
the two coupled cavities. $\mathrm{a_{P}-f_{P}}$ the reflected phase
of the two coupled cavities. The parameters are $r_{2}^{2}=0.958$,
$r_{0}^{2}=0.99$, $a_{1}=a_{2}=0$, $\phi_{1}=\phi_{2}$,
$n_{1}=n_{2}=1$, $\mathrm{a_{I}}$ and $\mathrm{a_{P}}$:
$r_{1}^{2}=0.999995$, $\mathrm{b_{I}}$ and $\mathrm{b_{P}}$:
$r_{1}^{2}=0.99995$, $\mathrm{c_{I}}$ and $\mathrm{c_{P}}$:
$r_{1}^{2}=0.9997$, $\mathrm{d_{I}}$ and $\mathrm{d_{P}}$:
$r_{1}^{2}=0.999$, $\mathrm{e_{I}}$ and $\mathrm{e_{P}}$:
$r_{1}^{2}=0.99$, $\mathrm{f_{I}}$ and $\mathrm{f_{P}}$:
$r_{1}^{2}=0.85$. }
\end{figure}

Fig. 1 plots the reflected amplitude and phase of the two coupled
cavities as a function of the cavity detuning for different coupling
strength by use of Eq. 3. The coupling strength (the reflectivity of
center mirror M1) increase with plot alphabet. Fig. 2 plots of the
quantum fluctuation spectra of the reflected field from the coupled
optical cavities for different coupling strength by use of Eqs. 7
and 8.

\begin{figure}[H]
\centering{
\includegraphics[width=3.5in]{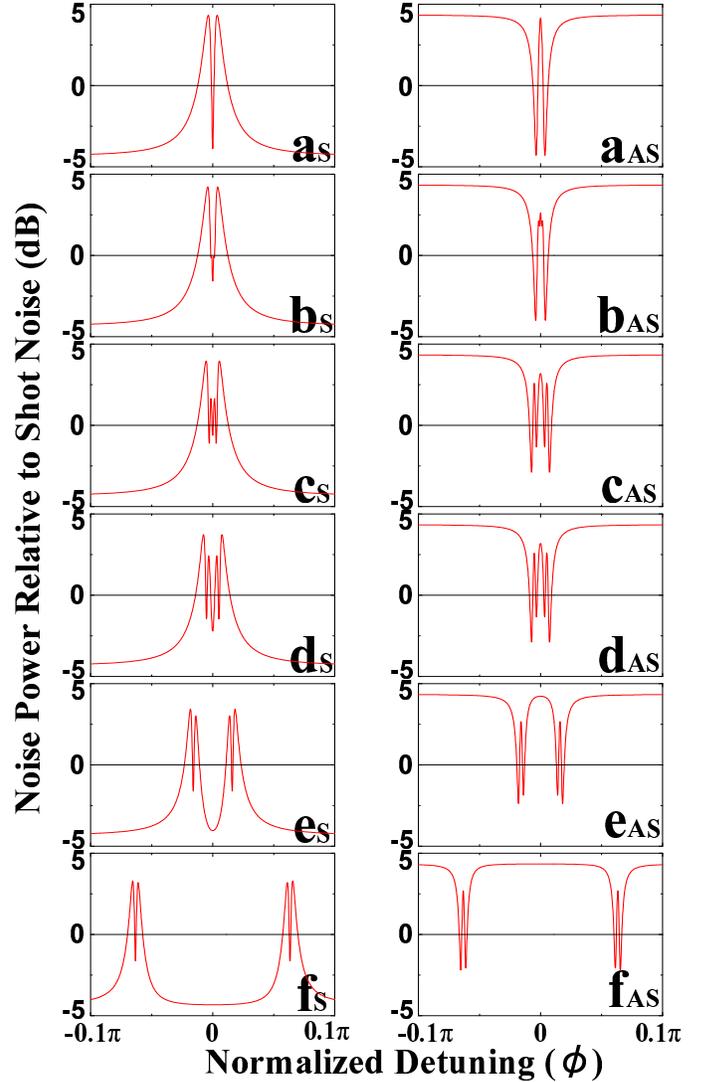}
} \vspace{0.1in} \caption{\label{Fig:epsart} (Color online).
Theoretical plots of the quantum fluctuation spectra of the
reflected field from the coupled optical cavities for different
coupling strength. $\mathrm{a_{S}-f_{S}}$ are the reflected
amplitude (as squeezing component) quadrature spectra for the two
coupled cavities. $\mathrm{a_{AS}-f_{AS}}$ are the reflected phase
(as antisqueezing component) quadrature spectra for the two coupled
cavities. The parameters are same as Fig. 1, $\Omega=0.0005c/2L$,
and $s=0.5$.}
\end{figure}

\end{document}